\documentclass[aps,prl,preprint,longbibliography]{revtex4-2}
\usepackage{graphicx}
\usepackage{bm}
\usepackage{color}
\begin{document}



\title{
Time-resolved synchronization analysis of stacked intrinsic Josephson junctions of a cuprate superconductor with frequency-modulated terahertz radiation spectra
}


\author{Masashi Miyamoto}
\author{Keisuke Mizoguchi}
\author{Ryota Kobayashi}
\author{Nozomi Yagyu}
\affiliation{
Department of Electronic Science and Engineering, Kyoto University, Kyotodaigaku Katsura, Nishikyo, Kyoto 615-8510, Japan
}
\author{Manabu Tsujimoto}
\affiliation{
National Institute of Advanced Industrial Science and Technology (AIST), Central 5, 1-1-1
Higashi, Tsukuba, Ibaraki 305-8565, Japan
}

\author{Itsuhiro Kakeya}
\email{kakeya@kuee.kyoto-u.ac.jp}
\thanks{Corresponding author}
\affiliation{
Department of Electronic Science and Engineering, Kyoto University, Kyotodaigaku Katsura, Nishikyo, Kyoto 615-8510, Japan
}


\date{\today}

\begin{abstract}
Terahertz radiation from $\text{Bi}_2\text{Sr}_2\text{CaCu}_2\text{O}_{8+\delta}$ intrinsic Josephson junctions (IJJs) provides an ideal platform to study the synchronization of a macroscopic quantum system. Here, we present a spectral analysis of a frequency-modulated Josephson plasma emitter coupled with patch antennas. In the unmodulated intensity distribution as a function of radiation frequency $I_{\mathrm{UM}}(\omega)$, we observe a double Gaussian peak structure. Crucially, double-peak spectra obtained at a constant bias voltage imply either a rapid temporal distribution of resonances or their simultaneous excitation, driven by the mutual electromagnetic coupling between the IJJ mesa and the antennas. At low modulation frequencies $f_m$, the spectra are well reproduced by the products of $I_{\mathrm{UM}}(\omega)$ and frequency combs, yielding a synchronized relaxation time $\tau_s \simeq 0.28\text{ ns}$. Incorporating $\tau_s$ quantitatively reproduces a drastic spectral transformation observed around $f_m \sim 1\text{ GHz}$, unveiling the sub-nanosecond non-equilibrium dynamics of coupled Josephson plasma.
\end{abstract}


\maketitle

\section{Introduction}

While intrinsic Josephson junctions (IJJs) ~\cite{Kleiner1992,Yurgens2000,Kleiner2019a} are well-established as intriguing model systems for interacting nonlinear oscillators~\cite{Jain1984,Wiesenfeld1998,Daniels2003}, a comprehensive understanding of their dynamic synchronization mechanisms, particularly under frequency modulation, remains elusive."
This is because two distinct interactions must be considered: inductive coupling arising from in-plane currents~\cite{Sakai:1993,Kleiner1994a} and capacitive coupling associated with the breaking of electrical neutrality at the superconducting electrodes~\cite{Koyama:1996,Machida1999}. These interactions are characterised by two distinct length scales: the penetration length $\lambda_ {ab} \sim 10^{-7}$ m and the Debye length $\mu \sim 10^{-10}$ m.
That is, it is thought that a nonlinear oscillator system, controllable by temperature or voltage, is realized where magnetic coupling acting across hundreds of Josephson junctions coexists with electrostatic coupling acting only between the nearest junctions~\cite{Machida2004}.

The characteristic of the IJJ system as a collective of Josephson plasma oscillations~\cite{Tachiki:1994,Kakeya1998} is also evident in the excitation method. An alternating current with a frequency proportional to the applied voltage is excited via the AC Josephson effect, and these two interactions enable its macroscopic observation as a collective phenomenon. One notable example of this achievement is the terahertz electromagnetic wave-emitting Josephson plasma emitter (JPE) of Bi$_2$Sr$_2$CaCu$_2$O$_{8+\delta}$ (Bi2212) ~\cite{Ozyuzer:2007,Kadowaki2008,Wang2010,Hu2010,Welp2013,Kakeya2016,Delfanazari2020}. The ability to relatively freely control the radiation frequency by voltage control~\cite{Benseman2011a,Tsujimoto2012,Kashiwagi2014,Borodyanskyi2017,Sun2023} makes it interesting not only for applications as a terahertz wave source~\cite{Tsujimoto2012b,Hao2015,Sun2017,Nakade2016,Asai2017,Elarabi2017,Saito2022,Elarabi2024} but also as a for novel phenomena in terahertz nonlinear optical phenomena and nonlinear oscillator many-body problems~\cite{Lin2008,Kakeya2012,Tsujimoto2014a,Benseman2019,Tsujimoto2020,Kobayashi2022,Kleiner2021}.
The collective phenomena in stacked IJJs can also be applied to a highly sensitive detector with a cascade detection~\cite{Cattaneo2025}.

While our previous work demonstrated frequency-modulated (FM) spectra and attributed the observed phenomena to a finite synchronization relaxation time~\cite{Miyamoto2024}, a precise, quantitative estimation of this relaxation time and a detailed discussion of its underlying mechanism were not presented. This quantitative gap forms the basis of the present study.
This study reports the spectral changes when the superimposed frequency is decrementally varied from $f_m=3$ GHz to 100 kHz, accurately estimates the relaxation time, and discusses the  mechanism for synchronising intrinsic Josephson junction systems.
To analyze the FM spectra quantitatively, frequency dependence of the radiation intensity under unmodulated (DC) bias voltages is discussed in detail.

\section{Methods}
\subsection{Experimental}
The JPE device was fabricated from a Bi2212 single crystal grown by the floating zone method and is identical to the device reported in Refs.~\cite{Miyamoto2024,Tsujimoto2021}, featuring triangular patch antennas as shown in the inset of Fig. \ref{fig:figure1} (a). A WF1968 function generator (NF Coop.) was used for applying DC and modulated bias up to 100 MHz, while a bias tee and a E4428C RF generator (Keysight Coop.) were employed for superimposing high-frequency signal to the bias up to 3.0 GHz. The bias circuitry was also identical to that of Ref.~\cite{Miyamoto2024}, though impedance was changed due to rewiring.
Biasing parameters are bias offset $V_{B0}$, superposing amplitude $P_m$ for the RF generator and $V_m$ for the function generator, and superposing frequency $f_m$.
Measured device voltage $V_\mathrm{dev}$ is slightly smaller than $V_{B0}$ but larger than the applied DC voltage $V_0$ because of voltage drops at the wirings and the contacts and the applied modulation amplitude $V_m$ depend on $f_m$ even for the equal $P_m$ as a result of $f_m$ dependent impedance of the biasing circuit $Z(f_m)$.
In this paper, the applied voltage to the stacking of $N$ IJJs is denoted as
\begin{equation}
V(t) =V_0+V_m \sin (2\pi f_m t),
\end{equation}
where $t$ is the time from the modulation origin $V=V_0$.

\subsection{Spectrum data processing}
The intensity of the detected electromagnetic waves, measured by the silicon bolometer of the Martin-Puplet-type spectrometer~\cite{Kakeya2013}, was plotted as a function of delay time $t_d$ . This plot was regarded as an interferogram $s(t_d)$.
The interferogram was numerically processed to a spectrum with the established method in fourier transform infrared (FTIR) spectroscopy.
Our original interferogram is asymmetric because $t_d$ (s) spans from $-0.02 \mathrm{(m)}/c_0$ to $+0.4 \mathrm{(m)}/c_0$ sec, where $c_0$ is the light velocity, thus we symmetrized the interferogram by taking "OR" aliasing with respect to the centerburst at $t_d=0$: symmetrized data at $|t_d|< 0.02/c_0$ sec and $|t_d|> 0.02/c_0$ sec are from $s(\pm t_d)$ average and $s(+t_d)$, respectively. Noted that phase correlation was performed to find the $t_d=0$.
Then, zeros are added to the both ends of the interferogram to be the length being $2^{17}= 131,072$ (original length: $~ 3\times 10^4$).
A consequent spectrum was obtained by applying a discrete FFT to the symmetrized and padded interferogram with the Hanning apodization.
This procedure makes the obtained spectra more precise and commensurable than our previous publications, where we have employed neither the symmetrization nor the zero filling.
More details are described in Supplementally Materials.

Figure \ref{fig:figure1} (a) shows the spectrum obtained by varying the DC bias $V_{B0}$, and Fig.~\ref{fig:figure1} (b) shows the variation of the integral intensity of the spectrum $I_{\mathrm{UM}}(\omega)$ with radiation frequency $\nu=\omega/2\pi$.
The radiation frequency is proportional to $V_0$ and its integration shows double-peak behavior.
The intensity decrease at the lower frequency end is very rapid, and no apparent radiation is found below 840 GHz.
$I_{\mathrm{UM}}(\omega)$ at 21.5 K excellently agrees with the sum of two Gaussian distributions centred at $\nu_\mathrm{peak1}=849$ GHz with a full-width at half maximum (FWHM) of  5.0 GHz and centered at $\nu_\mathrm{peak2}=866$ GHz with a FWHM of 30 GHz.

\subsection{FM spectrum with $\omega$-dependent carrier amplitude}
The temporal variation of the ideal FM signal electric field is given by
\begin{equation}\label{eq:e-field}
u_{\mathrm{FM}}(t) = A \cos\left[\omega_c t + m_f \sin(2\pi f_m t)\right]
= A \sum_{n=-\infty}^{\infty} J_n(m_f) \cos\left[(\omega_c + n 2\pi f_m)t + \phi_0\right],
\end{equation}
where $A$ is the amplitude of the carrier-wave with angular frequency of $\omega_c=2\pi \nu_c$, $m_f$ is the modulation index, $J_n(x)$ is the $n$-th order Bessel function of the first kind, and $\phi_0$ is a constant.
Since the radiation amplitude of the JPE exhibits frequency dependence, we employed $A(\omega)$ instead of $A$.
When $A(\omega)$  depends solely on $\omega$, the intensity spectrum becomes the product of the radiation intensity frequency dependence $I_{\mathrm{UM}}(\omega) = \frac{1}{2} A^2(\omega)$ and the FM comb, resulting in
\begin{equation}\label{eq:spectrum1}
I_{\mathrm{FM}}(\omega)
= \langle u_{\mathrm{FM}}^2(t) \rangle_t
= I_{\mathrm{UM}}(\omega) \sum_{n=-\infty}^{\infty} J_n^2(m_f) \, \delta(\omega - \omega_c -  2\pi n f_m),
\end{equation}
which is an intuitively understandable expression.

In the FM-JPE, when we apply a bias of $V_0+V_m\cos2\pi f_m t$ to the stacked $N$ IJJs, the carrier-wave frequency $\nu_c=2eV_0/h N$ and the maximum frequency deviation $\nu_d=m_f  f_m=2eV_m/h N $ are obtained, where $e$ and $h$ are the elementary charge and the planck constant, respectively~\cite{Miyamoto2024}. It is noted that $V_m$ is given by output power $P_m$ of the RF generator and impedance of the bias circuit $Z(f_m)$.
The spectrum near the lower edge of the radiation frequency region disagrees with Eq. (\ref{eq:spectrum1}).
For $f_m=3$ GHz, even at $\omega$ where $I_{\mathrm{UM}}(\omega) \simeq 0$, i.e., $A(\nu)\simeq 0$ holds, not only is a peak clearly visible equivalent to that at $\omega$ with much larger $A(\omega) $, but the strong frequency dependence of $I_{\mathrm{UM}}(\omega)$, as seen in Figure \ref{fig:figure1} (b), is not reflected to the spectra shown in Fig. 3 (a) of Ref. ~\cite{Miyamoto2024}. To resolve this contradiction, we discuss the synchronisation relaxation time $\tau_s$ of Josephson plasma oscillations excited by IJJ.

\subsection{Determination of bandwidth}
To quantitatively evaluate changes in FM spectrum, we focus on the spectral bandwidth and the integrated intensity.
To determine the bandwidth and center frequency $\nu_c$ of the experimentally obtained modulation spectrum, the power spectrum was numerically integrated as follows: over approximately 200 GHz range centred on the frequency band where the spectrum appears.
In the common definition of communication technology, spectrum integration over the bandwidth covers 98 \% of the radiated power of interest.
In the present study, we determine frequencies giving 5 and 95 \% values of the whole integration as lower and upper limits of a band, respectively, and their difference was defined as the bandwidth $B$.
This criterion facilitates the extraction the variation of the spectrum as a function of $f_m$ by avoiding extrinsic floor noise in the spectrum.

\section{Results and discussion}
\subsection{Double Resonance Structure in the Frequency Dependence of the unmodulated Spectral Intensity}
The double peak structure of $I_\mathrm{UM}(\omega)$ is one of the central issue addressed in this study.
The primary factors determining JPE radiation intensity are three: (1) the maximum Josephson current density in the IJJs under radiation conditions $J_0$; (2) the number of junctions synchronously oscillating $N_s$ within the approximately 1000 IJJs; and (3) the radiation efficiency determined by the structure of the mesa, antenna, and electrodes~\cite{Tachiki2011,kobayashi2025}.  Among these, $J_0$ is dominated by the Bi2212 chemical composition and measurement conditions such as temperature.
$N_s$ is partly determined by the frequency characteristics of the antenna radiation circuit. The frequency characteristics of the device enable it to possess a double-resonance structure.
For the resonant structures of this device, there is a patch antenna structure in addition to the mesa structure.

\begin{figure}[htbp]
  \centering
  \includegraphics[width=\linewidth]{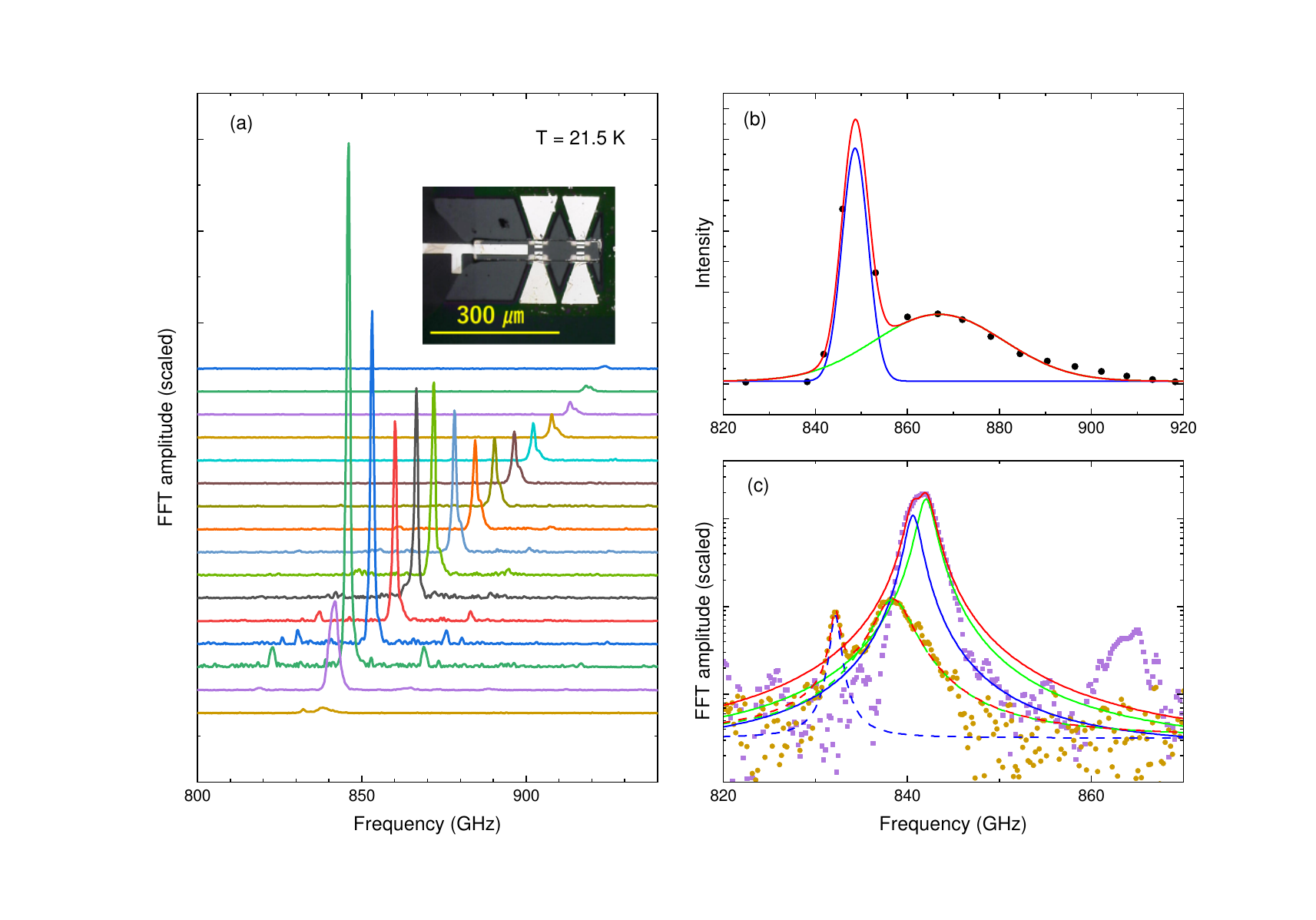}
  \caption{
  (a) Mesa voltage dependence of the radiation spectrum for $V_m=0$ at bath temperature $T$ of 21.5 K. The difference from Ref.~\cite{Miyamoto2024} lies in the frequency resolution and the data processing. Small satellites $\pm 20$ GHz apart from the main peak are presumably artifacts depending on the apodization for discrete Fourier transformations. Inset is the microscope photograph of the device.
  (b) Integrated spectrum intensity as a function of radiation frequency $I_\mathrm{UM}(\omega)$. Solid lines are Gaussian fittings with center frequencies are 849 GHz and 866 GHz and their sum.
  (c) Double Lorentz peak fitting for the spectrum centered at 841 GHz (purple squares and solid curves). The spectrum is decomposed to two Lorentz peaks with center frequencies of 840 and 842 GHz and linewidths of 1.7 and 1.9 GHz, respectively.
  Spectrum with apparently separated peaks (dark-yellow circles) and its fittings (broken curves) are also depicted.
  }
  \label{fig:figure1}
\end{figure}

In Ref.~\cite{Tsujimoto2021}, numerical calculations of a JPE device with patch antennas of the identical design are reported.
The real part of the antenna impedance is peaked at 750 GHz with an FWHM of 50 GHz. This is reminiscent of the broader peak of $I_{\mathrm{UM}}(\omega)$.
This can be interpreted that the patch antennas are attributed to the broad intensity distribution with the center at 866 GHz in this study.
With using the $TM_{m,n}$ mode cavity resonance formula of the mesa structure~\cite{Kashiwagi2011}, $f_{m,n}=c_0/2n_r\sqrt{(m/L)^2+(n/W)^2}$, $f_{1,3}=866$ GHz is the closest mode to our result for the refractive index $n_r=4$, where
$L=48 \mu$m and $W=300 \mu$m are actual lengths of short and long edges of the mesa, respectively.
In the simple cavity resonance picture, $TM_{1,3}$ mode has magnetic antinodes at $y=0, W/3, 2W/3, W$ along the long edge of the mesa.
According to the device photograph (inset of Fig. \ref{fig:figure1} (a)), the antennas are connected to the mesa at $y\simeq$ 50 and 150 $\mu$m from a short edge, around which the magnetic nodes of the $TM_{1,3}$ mode exist.

In the device without an antenna fabricated with the similar fabrication process and geometries as shown in the \emph{Supplementary Materials}, we found that the FWHM of $I_{\mathrm{UM}}(\omega)$ of a single mode is approximately 22 GHz.
This is closer to the broader peak shown in Fig. \ref{eq:spectrum1} (b).
Furthermore, another radiation region was also found at proximate frequencies ($\sim 500$ GHz) with different bias voltages.
This result provides important insights for designing JPE device, as the addition of the patch antenna results in drastic change in radiation properties of IJJ mesas.
To highlight the quantitative effect of the patch antenna on radiation, a three-dimensional simulation of the electromagnetic field is necessary, as the device’s structure is too complex to analytically determine the resonance modes responsible for the radiation.
This is beyond the scope of this study and will be published in the future.

Careful observation of the spectrum near the lower edge of the radiation frequency range, around 840 GHz, reveals not only a broadened linewidth but also what appears to be multiple peaks. Decomposing this spectrum into respective peaks yielded two Lorentz functions with linewidths of $\Delta \nu =$1.7 and 1.9 GHz as shown in Fig. \ref{fig:figure1}(c). For slightly smaller $V_\mathrm{dev}$ values, more separated and broadened spectrum and its fitting are also shown in Fig. \ref{fig:figure1}(c).
The Lorentz fitting linewidths for more intensive spectra for $850<\nu_e<900$ GHz are distributed 1.0$ < \Delta \nu<$ 1.5 GHz, suggesting the spectrum with lower $\Delta\nu$ exhibits closer characteristics. Thus, instances where multiple peaks are simultaneously observed in JPE have already been reported~\cite{Zhang2019}, where it is argued that the radiation frequency is distributed temporally or spatially.
The finding of this double-peak feature is a result of improved frequency resolution attributed to the long $t_d$ scan and the introduction of the standard interpolation (symmetrization and padding) method.

For radiation frequencies distributed temporally due to transient between zero and finite resistance of any included IJJs, discontinuous changes would be observed in the interferogram if the timescale were several seconds or longer; however, none were detected in this study. Furthermore, the standard deviation of $V_\mathrm{dev} \simeq V_0$ remained unchanged compared to the spectral measurement at $\Delta \nu \simeq$ 1.2 GHz, indicating that fluctuations in the bias voltage were not the cause of the broadening. Considering the possibility that the radiation frequency is spontaneously modulated rapidly, similar to the results obtained with an FM spectrum from high-frequency superimposition on the bias, it can also be hypothesised that the radiation frequency or intensity spontaneously modulated at frequencies in the MHz order or higher, resulting in the double-peak spectrum.

The spatial distribution of radiation frequencies should be considered in relation to the fact that radiations is obtained from the resonant modes of both the mesa and the antenna parts of the device discussed earlier. Since the radiation linewidths $\Delta \nu$, apart from the peak at 840 GHz currently under consideration, are all approximately 1.0 GHz, it is thought that $\Delta \nu$ is determined not by the $Q$-factor of the resonant structure, but by the temporal and spatial inhomogeneity of the bias along the stacked IJJs.
The multiple Gaussian peak fits shown in Fig. \ref{fig:figure1}(b) reveal the dominant radiation modes at these frequencies:
the $843 < \nu < 855$ GHz region is inferred to be related to the mesa structure, while the $855 < \nu < 910$ GHz region is suggested to be related to the antennas, by following the interpretation of Ref.~\cite{Tsujimoto2021}.
The IJJ system selects one of these modes. However, at $\nu=842$ GHz, both resonance modes are close in intensity and frequency. Consequently, it is considered that transitions occur between these two resonance modes on a short time scale, resulting in a spectrum comprising multiple peaks despite the absence of superimposed bias. Therefore, in spectra exhibiting multiple closely spaced peaks, it is thought that bifurcations occur on short timescales between different radiation modes spatially distributed within the device.

\subsection{Frequency modulated spectrum}
\subsubsection{$f_m$ dependence of the FM spectrum}
\begin{figure}[htbp]
  \centering
  \includegraphics[width=\linewidth]{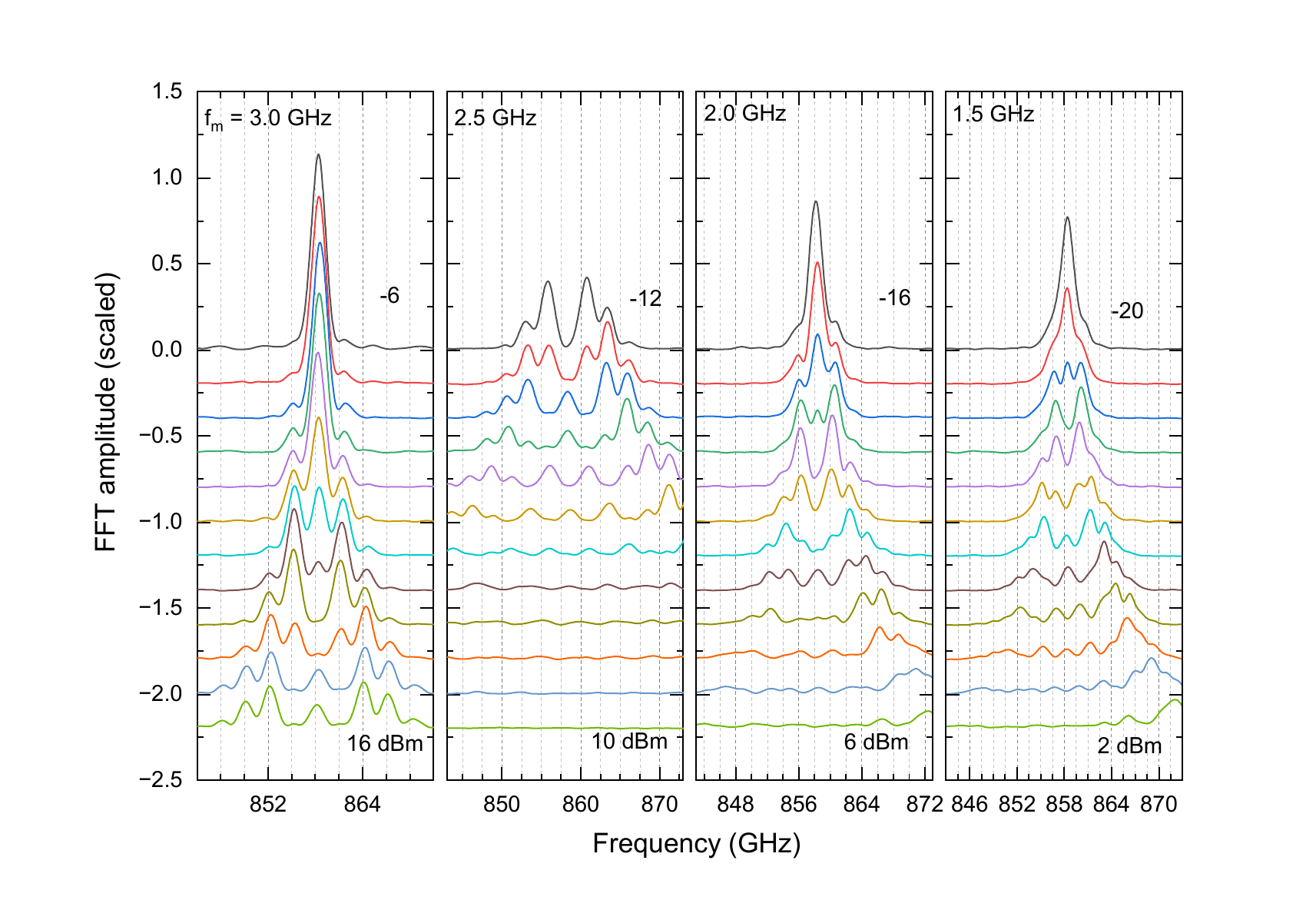}
  \caption{Frequency modulated spectrum when superimposed frequency $f_m$ is varied from 3.00 GHz to 1.50 GHz with every 2 dBm different output power $P_m$ of superimposed signal.
  Expanding sidebands are more intensive for lower frequency (left) side at $f_m=3.0$ GHz, while for higher frequency (right) side in  $f_m<2.5$ GHz.
 The spacing of subgrids corresponds to each $f_m$, thus equiseparated peaks are clearly found.
  }
  \label{fig:figure2}
\end{figure}

Figure \ref{fig:figure2} shows the spectra centered at $\nu=858 $ GHz, slightly higher than the sharp intensity peak, with superimposed frequencies $f_m$ from 3.0 to 1.5 GHz. It can be confirmed that the spacing between the comb teeth agrees with $f_m$.
With increasing $P_m \propto V_m^2$ at an $f_m$, other oeaks appear at both ends of the spectra, with following Eq. (\ref{eq:spectrum1}).
In this range of the modulation bandwidth, the extents of the FM sidebands, approximately up to 15 GHz, peaks at higher side of the spectra are more remarkable than those at lower frequency side for $f_m<$2.5 GHz. This trend is similar to that of the broad peak in $I_{UM}(\omega)$. However, at $f_m=$3.0 GHz,  the lower sidebands at 855 GHz are more remarkable than the higher sidebands, which is attributed to the delay in teh radiation intensity with respect to the instantaneous radiation frequency. This feature is another central issue of this study.
It is also found that the spectral bandwidth at the same $P_m$ depends on $f_m$. This is attributed to the $f_m$ dependence of impedance $Z(f_m)$ of the biasing circuit, which includes not only coaxial cable and wires but also silver paste for bonding wires. The superimposed signal power $P_m$ is generally more dissipated on the way to the JPE at higher $f_m$ within this range.

%

\subsubsection{Spectral cutoff with unmodulated intensity distribution}
\begin{figure}[htbp]
  \centering
  \includegraphics[width=\linewidth]{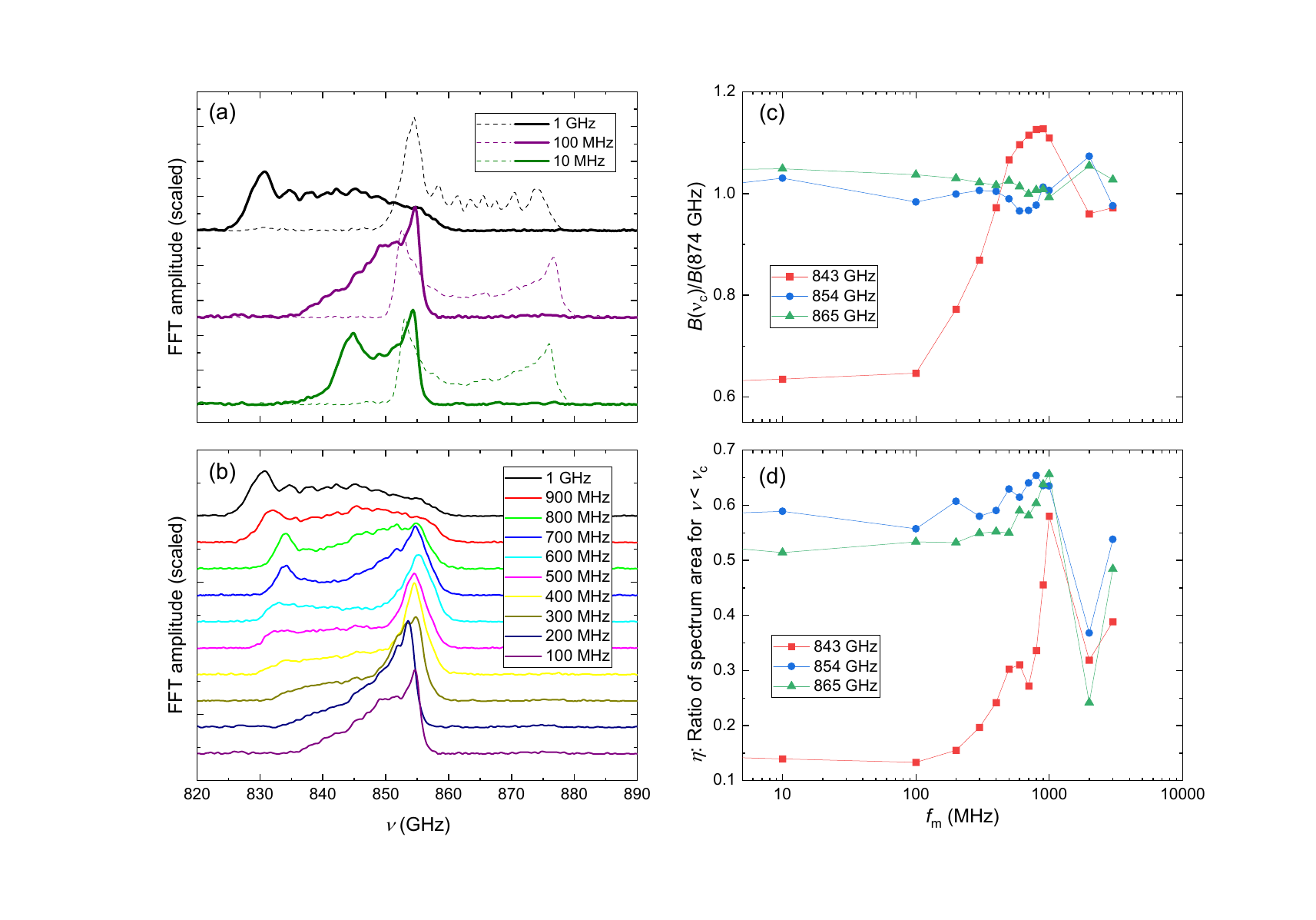}
   \caption{
  (a) FM spectra centred at 865 GHz with $V_\mathrm{B0}=$ 2.45 V (broken curves), where $I_\mathrm{UM}(\nu)$ has the global maximum and spectra at 843 GHz with $V_\mathrm{B0}=$ 2.36 V (solid curves), near the lower edge of the unmodulated radiation frequency range.
  (b) Detailed modulation frequency $f_m$ evolution of FM spectra. Superimposed signal power $P_m$ is adjusted so that the maximum frequency deviation $\nu_d$ to the higher frequency side are approximately equalized.
  (c) Comparison of bandwidth of the spectrum as a function of $f_m$ for various center frequencies $\nu_c$ determined with $V_0$. Data are normalized by values for $\nu_c=874$ GHz and $V_\mathrm{B0}=$ ? V. The bandwidth for the $\nu_c=$ 843 (red) spectrum linearly increase with $f_m$ between 100 and 600 MHz.
  (d) $\eta(f_m,\nu_c)$ for various $\nu_c$s. Variation for $f_m \ge $1 GHz is similar among all $\nu_c$s.
  }
  \label{fig:figure3}
\end{figure}

The broken curves in Fig. \ref{fig:figure3} (a) correspond FM spectra with $\nu_c=865$ GHz, near the center of the unmodulated radiation frequency range, with approximately 30 GHz bandwidth at the superimposed frequencies $f_m=1$ GHz, 100 MHz, and 10 MHz. Here, $V_{B0}$ was fixed at 2.42 V, and $P_m$ was adjusted to equalize the spectral bandwidth, those are $P_m= -16$ dBm, $-20$ dBm, and $-20$ dBm, respectively. The device voltage $V_\mathrm{dev}$ measured during interferogram acquisition was $V_\mathrm{dev}=2.15 \pm$ 0.035 V for all $f_m$. Since this was independent of $f_m$, it is evident that the device and measurement system inherently possess voltage fluctuations of $\pm$ 1.6 \%. At $f_m$ exceeding the spectrometer resolution ($ \simeq 0.8$ GHz), comb-like features are observed, but these are absent at lower $f_m$, making it difficult to distinguish the peaks appearing at the $f_m$ intervals. All spectra share the common feature of having a prominent peak at the band edges, with the lower-frequency edge near $\nu=$ 853 GHz exhibiting a higher peak than at the higher edge. This asymmetry presumably originates from $I_{\mathrm{UM}}(\omega)$.
Meanwhile, the thick curves in Fig. \ref{fig:figure3}(a) present the spectra under equivalent conditions centerd around $\nu_c=$843 GHz (with corresponding $V_{B0}$), near the lower edge of the unmodulated radiation frequency range. The spectral characteristics differ significantly for each $f_m$. First, at  $f_m =$1 GHz, the spectrum is generally weaker overall, and while the comb teeth are less prominent, they remain visible at 1 GHz intervals, qualitatively similar to the case of $\nu_c$=865 GHz, as shown as a thin curve. At 100 MHz, not only do the comb teeth disappear, but the left half of the spectrum appears to be missing, with intensity gradually increasing towards higher frequencies from 840 GHz. At 10 MHz, another peak was found at a frequency of 845 GHz, yielding an FM spectrum with a global feature different from both 100 MHz and 1 GHz.

We now examine the spectra with the superimposed frequency $f_m$ varied  between 1 GHz and 100 MHz as shown in Figure \ref{fig:figure3} (b), separating the regions below and above 840 GHz. On the low-frequency side ($\nu<$ 840 GHz), as the superimposed frequency $f_m$ decreases, the spectrum intensity diminishes and shifts towards the high-frequency side, vanishing at $f_m=$ 600 MHz. At even lower $f_m$s, it gradually decreases, with virtually no integrated intensity below 840 GHz at $f_m=$ 100 MHz. The high-frequency side exhibits broadly the opposite behaviour. The peak at the high-frequency band edge becomes more prominent as the superimposed frequency decreased. The peak height reaching its maximum at 200 MHz is thought to be due to the modulation bandwidth being slightly smaller and the upper-side of the maximum-deviated frequency being close to the sharp maximum in $I_{\mathrm{UM}}(\omega)$ at $\nu_\mathrm{peak1}$=850 GHz of the unmodulated spectral intensity $I_\mathrm{UM}(\omega)$.

\subsubsection{Quantitative analysis of the spectrum cutoff}

A careful observation of the FM spectrum revealed that as the superimposed frequency increases from 100 MHz, the spectral intensity below 840 GHz increased.
Therefore, the spectral bandwidth $B$ at various $\nu_c$ are plotted as a function of $f_m$ in Fig. \ref{fig:figure3} (c). Here, the values are normalized by data for $\nu_c=874$ GHz because of the difference in the actual $V_m$ for different $f_m$s and resulting in variation in maximum frequency deviation $\nu_d$. With increasing $f_m$ the intensity at $\nu_c=$ 843 GHz starts to increase at 100 MHz, then approaches to the maximum at 900 MHz whereas data for other $\nu_c$s does not depend on $f_m$.
This analysis highlights the change in the extent of the FM spectra as a function of $f_m$.

For simplicity, we consider that the radiation intensity has a step-function $\nu$-dependence:
$I(\nu)=1$ for $\nu_L<\nu<\nu_U$ and $=0$ for others.
Assuming that $\nu_c$ corresponds to $\nu_L$, $B(\nu_c)=\nu_d$ for $f_m \tau_s=0$.
With increasing $f_m \tau_s$, $B(\nu_c)$ increases as $\nu_d(1+\sin 2\pi f_m \tau_s)$ up to $f_m \tau_s=0.25$ then saturates because the radiation survives for $\tau_s$ even when the instantaneous frequency $\nu < \nu_L$, which does not delay from the modulated bias voltage.
In our experiment, 843 GHz is considered to be slightly higher than $\nu_L$ thus $B(\mathrm{843 GHz})=\nu_d(1+\delta)$ with 0 $<\delta \ll$ 1.
It is noted that  $B(\mathrm{843 GHz})$ does not depend on $f_m$ when $f_m \tau_s < \delta/ 2\pi$.
Considering $B(\mathrm{874 GHz})=2 \nu_d$, with increasing $f_m \tau_s$, $B(\mathrm{843 GHz})/B(\mathrm{874 GHz})$ varies from 0.5+$\delta$ to 1 and saturates at $f_m=0.9$ GHz.
Thus, based on this simplified model, a synchronized lifetime $\tau_s \simeq $ 0.28 nsec is estimated for this device.
We expect that $\tau_s$ is determined by the strength of the inter-IJJ coupling and depends on the device temperature, but its evolutions are remained for further research.

As another evaluation procedure, spectral intensity ratio for $\nu < \nu_c$ was calculated as follows:
Figure \ref{fig:figure3} (d) shows the spectral intensity ratio below the carrier frequency $\nu_c$ to the whole spectrum intensity as a function of $f_m$
\begin{equation}
\eta(f_m,\nu_c)=\frac{\int_{\nu_-}^{\nu_c} I_{\mathrm{FM}}(\nu) d\nu}{\int_{\nu_-}^{\nu_+} I_{\mathrm{FM}}(\nu) d\nu},
\end{equation}
where $\nu_-$ and $\nu_+$ are the lower and upper limit frequencies of the numerical integration of the spectrum.
The value of $\eta$ is expected to represent the agreement between experimental spectra and Eq. (\ref{eq:spectrum1}).
With increasing $f_m$, $\eta$ starts to increase at $f_m=$ 0.2 GHz and reaches to a maximum at $f_m=$ 1 GHz, where we employed $\nu_{\pm}=\nu_c\pm$ 15 GHz, which are common among different $f_m$s.
This analysis illuminates the distribution change of the Josephson oscillators inside the mesa structure with varying $f_m$ more clearly than the comparison of the bandwidth.
We notice that as $f_m$, $\eta(f_m,843 \mathrm{GHz})$ approaches to 0.5 and  $\eta(f_m,\nu_c)$ at higher $\nu_c$s and never exceeds them.
This can be attributed to the conservation of the  total radiation intensity within the modulation range.

Detailed observation of FM spectra facilitates the time-resolved analysis of dynamical and non-linear phenomena in intrinsic Josephson junction systems. In this context, the time-resolved characterization of mode-locking in quantum cascade lasers using frequency combs~\cite{Abajyan2025} represents a pioneering endeavor that paves the way for such FM-based time-resolved studies.


\subsubsection{Spectral intensity frequency dependence and relaxation time introduction}


Here, we attempt to show that the measured FM spectrum changes to less expected by Eq. (\ref{eq:spectrum1}) at higher $f_m$ and it is necessity to introduce a finite relaxation time $\tau_s$ for the equilibrium radiation intensity.
We obtained an FM spectrum and estimated its bandwidth from the integration.
Then, the modulation index is derived from the bandwidth $m_f=B/2 f_m$, and a set of delta functions is created according to Eq. (\ref{eq:spectrum1}).
At this stage,  $I_\mathrm{UM}(\omega)$ is obtained from the unmodulated bias experiment, the fitted red curve in Fig. \ref{fig:figure1} (b) is used.

\begin{figure}[htbp]
  \centering
  \includegraphics[width=0.7\linewidth]{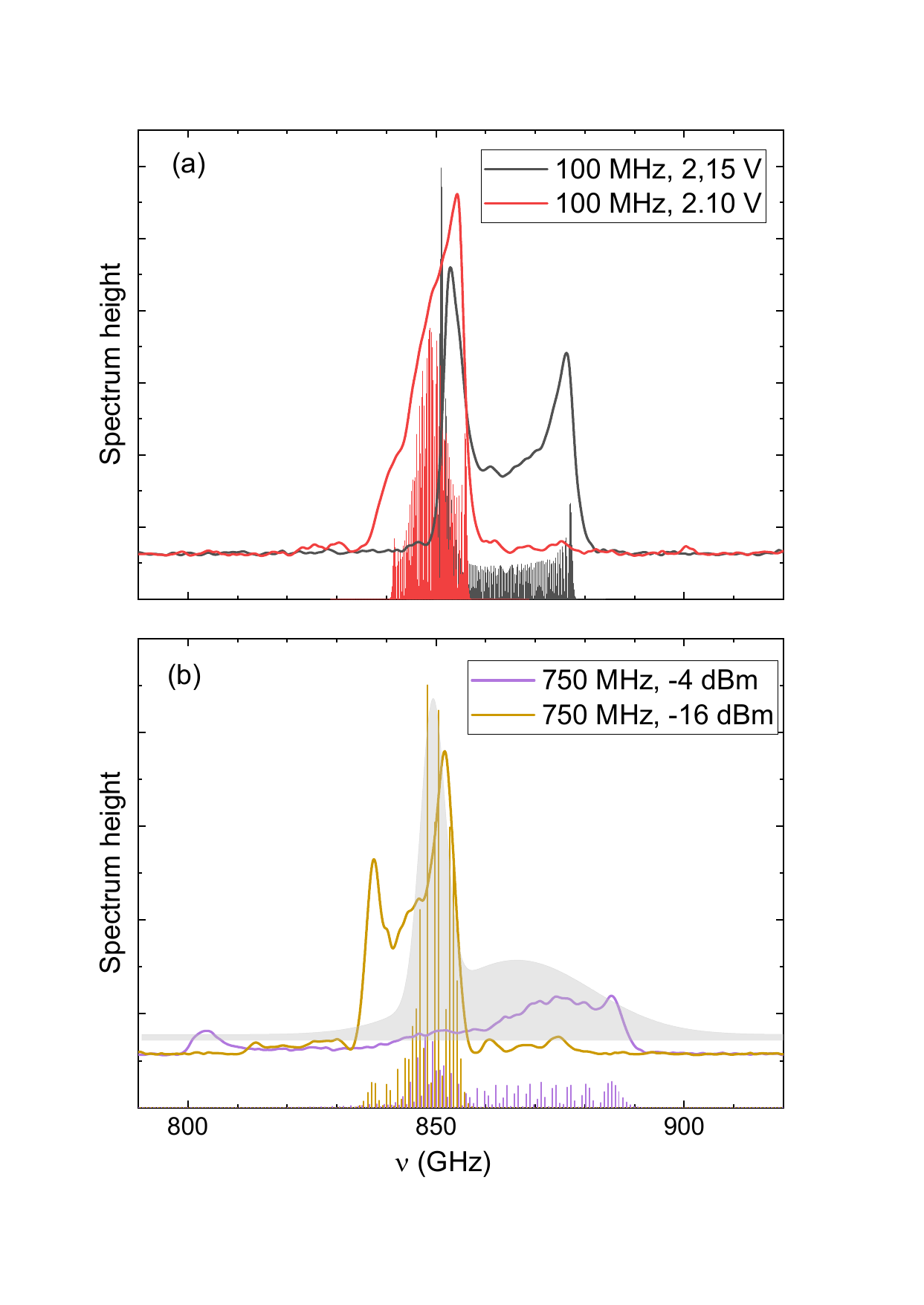}
    \caption{(a) Spectra at $f_m=$ 100 MHz,  $P_m=-20$ dBm, $V_\mathrm{B0}=$ 2.42 (black), 2.36 V (red). The black (red) bar graphs represent numerically reproduced spectra using center frequency 864.1 (848.7) GHz and a bandwidth 26.75 (14.95) GHz. (b) Spectra at $f_m=$ 750 MHz, signal outputs of $P_m=-4$ (purple) and $-16$ (dark yellow) dBm, and $V_\mathrm{B0}=$  2.40 V ($V_\mathrm{dev}=$2.1106 V). The bar graphs show the reproduced spectrum using $\nu_c=$ 845.6, 845.2 GHz and $B=$ 84.20, 18.9 GHz respectively.
    The gray shadow represents radiation the intensity distribution of the unmodulated bias $I_\mathrm{UM}(\omega)$.
    }
  \label{fig:figure5}
\end{figure}

Figure \ref{fig:figure5} (a) shows experimental (curves) and calculated (bars) spectra at two $\nu_c$s with $f_m=100$ MHz and $P_m= -20$  dBm, where the FM spectrum was cut at $I_\mathrm{UM}(\omega)$.  For both $\nu_c$s, the envelopes match perfectly. The comb-like variation found in the calculations is not  thought to be reproduced because of the constraints imposed by the experimental frequency resolution $\simeq 0.8$ GHz. However, as $f_m$ is increased, unexpected peaks at certain frequencies are observed, and the bandwidth differed significantly from the calculations. This is examined this in detail in the following:
In the FM spectrum with a sufficiently large $m_f$, peaks appear at the band edges.
As shown in Figure \ref{fig:figure5}(b), the band edge peaks in the $P_m=-4$ dBm spectrum at $f_m=$ 0.75 GHz corresponded to this. However, the peak at the lower band edge was clearly observed despite the frequency region being $I_\mathrm{UM}(\omega)\simeq 0$.
For these spectra, the superimposed frequency $f_m$ was less than the spectrometer's resolution; therefore comb-like peaks are not clearly recognized. Furthermore, as $f_m$ increases, comb-like peaks become apparent, but the height ratio appears unrelated to $I_\mathrm{UM}(\omega)$ (gray area). In the region $f_m$ above 2 GHz, this tendency is particularly pronounced in spectra with modulation bandwidths exceeding 50 GHz.


\subsubsection{Numerical reproduction of the change in spectrum symmetry according to the relaxation time for synchronization}

To reproduce the change in the height of the edge peaks as a function of $f_m$, we consider a simple model for the radiation intensity.
When considering the synchronization relaxation time $\tau_s$ associated with the temporal variation in $\omega=2\pi\nu$, we assume that
the amplitude of the radiation is delayed $\tau_s$ from the instantaneous angular frequency $\omega(t)$ given by the instantaneous voltage $V(t)$ applied.
The amplitude $A(\omega^{\prime})$ of the synchronised result is then given by
\begin{equation}\label{eq:freq_delay}
\omega^{\prime}(t) = \omega(t - \tau_s) = \omega_c + m_f \cos\left[2\pi f_m (t - \tau_s)\right],
\end{equation}
where $\omega(t)$ has no delay from $V(t)$.
From this, the electric field amplitude considering $\tau_s$ is obtained as follows: The FM electric field can be expressed as
\begin{equation}\label{eq:delayed_e-field}
u^{\prime}_{\mathrm{FM}}(t) = A(\omega^{\prime}) \cos \left( \int_0^t \omega^{\prime}(t') dt' \right) = A(\omega^{\prime}) \sum_{n=-\infty}^{\infty} J_n(m_f) \cos\left(2\pi(\nu_c + n  f_m)t + \phi_0'\right),
\end{equation}
and the spectrum is obtained from $\langle u^{\prime 2}_\mathrm{FM} (t) \rangle$, though an analytical expression is untrivial.

\begin{figure}[htbp]
  \centering
  \includegraphics[width=0.8\linewidth]{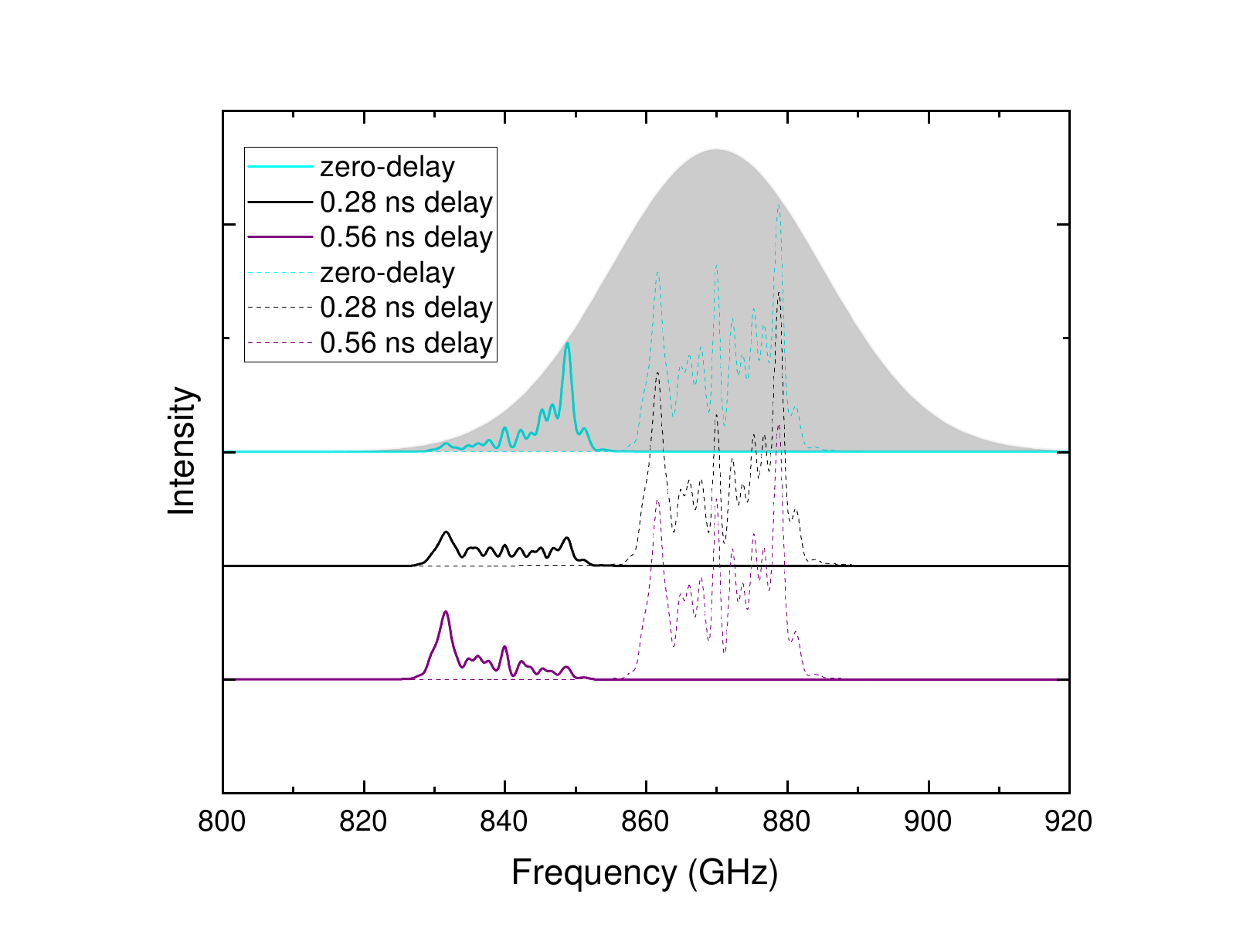}
  \caption{Numerically obtained FM spectra at $f_m = 1$ GHz $\nu_c =$ 840 and 870 GHz (solid and broken curves, respectively), and $\tau_s = 0$, 0.28, 0.56 ns (cyan, black, purple respectively) in comparison with the assumed gaussian intensity distribution peaked at 870 GHz with 15 GHz standard deviation (grey shadow).
 The comb teeth shift towards lower frequencies with changes in relaxation time for $\nu_c =$ 840 GHz, exhibiting symmetry opposite to the underlying resonance distribution.
 }
  \label{fig:figure6}
\end{figure}

Therefore, assuming a constant $\tau_s$, we calculated the time evolution of $u^{\prime 2}_{\mathrm{FM}}(t)$ corresponding to the experimental interferogram, performed a Fourier transform, and presented the resulting spectrum in Figure \ref{fig:figure6}. Here, a virtual unmodulated radiation intensity distribution $\tilde{I}_\mathrm{UM}(\omega)$ is assumed to follow a simple Gaussian distribution with the peak frequency of 870 GHz and the standard deviation of 15 GHz (the FWHM of 35.3 GHz). When $\omega_c$ coincides with the peak frequency of $\tilde{I}_\mathrm{UM}(\omega)$, it exhibits a comb-like spectrum independent of $\tau_s$. Conversely, when $\omega_c$ is lower than the frequency at the maximum of $\tilde{I}_\mathrm{UM}(\omega)$, $\tau_s=0$ yields a comb-spectrum with higher peaks at $\omega > \omega_c$ according to $\tilde{I}_\mathrm{UM}(\omega)$, while $\tau_s > 0.28$ nsec in the present result,  $\simeq f_m^{-1}/4$, exhibits a spectrum with higher peak at $\omega < \omega_c$, $\tilde{I_\mathrm{UM}}(\omega)$ with a smaller value.
This result is consistent with the spectrum at $f_m=1$ GHz as shown in Fig. \ref{fig:figure3}(a,b), where the peak on the low-frequency edge is more pronounced than at the other edge near $f_m$=1 GHz, differing from the other $f_m$s. This indicates that while the radiation frequency responds without delay to voltage modulation, the intensity requires a finite $\tau_s$ for IJJ coupling.

This synchronous relaxation time corresponds, in other words, to the lifetime of the macroscopic plasmon excited within the mesa structure. Furthermore, this plasmon lifetime differs from the microscopic lifetime of the Josephson plasmon, which provides the linewidth observed in the millimeter-wave range of Josephson plasma resonance in Bi2212 single crystals~\cite{Gaifullin1999,Kakeya2005}.

On the other hand, in this discussion where $\tau_s$ is fixed, increasing only shifts the prominent peak. It is readily apparent that a symmetric spectrum cannot be obtained at the edge of the unmodulated radiation region, as observed in the $f_m=$ 3 GHz experiment~\cite{Miyamoto2024}.
To achieve this reproduction, it is necessary to introduce a relaxation rate distribution, such as a Poisson distribution of the Drude model, which demands advanced numerical calculations that are beyond the scope of the present study.

\section{Summary}
The detailed radiation spectra of frequency-modulationed (FM) Josephson plasma emitter in terahertz (THz) frequency range have been investigated.
The unmodulated (DC) bias radiation provides the overlapping double-gaussian radiation intensity distribution, $I_{UM}(\omega)$, as a function of the DC voltage.
The radiation spectra at a specific range in the radiation frequency exhibit double-peaks.
The origin of these may be attributed bistable resonance mode yielded by the triangular patch antenna structures of the device.
Further circuit~\cite{kobayashi2025} and three-dimensional numerical simulations of the radiating electric field would be necessary to fully elucidate the origin.

The synchronization relaxation time
$\tau_s$ of 0.28 nsec at 21.5 K was estimated by analysing the frequency-modulated spectra at the edge of the double-gaussian radiation region.
The spectra were expanded beyond the radiation region from 0.2 to 1 GHz, which are quantified by the FM bandwidth $B(\nu_c,f_m)$ and the intensity ratio of the lower half to the total radiation power.
A simple numerical simulation of radiation spectra with $\tau_s$ into considered for the radiating intensity (but no delay in the frequency) partially reproduces the evolution of FM-spectra for $f_m \sim 1 $ GHz.
Time-resolved analyses can be extracted from FM spectra by using frequency dependent radiation intensity.


%


%



\clearpage

\begin{acknowledgments}
This work was supported by the Japan Society for the Promotion of Science (JSPS) KAKENHI (Grant Nos. JP23K17747 and JP24K000946).
\end{acknowledgments}

\bibliography{library}

\end{document}